\documentclass[%
 reprint,
 amsmath,amssymb,
 aip,
]{revtex4-2}

\usepackage{graphicx}
\usepackage{dcolumn}
\usepackage{bm}
\usepackage[caption=false,listofformat=subsimple,labelformat=simple]{subfig}

\usepackage{hyperref}
\begin{document}

\preprint{APS/123-QED}

\title{Viscoelastic response of impact process on dense suspensions}

\author{Pradipto}
  \email{pradipto@yukawa.kyoto-u.ac.jp}
\author{Hisao Hayakawa}%

\affiliation{Yukawa Institute for Theoretical Physics, Kyoto University, Kitashirakawaoiwake-cho, Sakyo-ku, Kyoto 606-8502, Japan}%

\date{\today}

\begin{abstract}
We numerically study impact processes on dense suspensions using the lattice Boltzmann method to elucidate the connection between the elastic rebound of an impactor and relations among the impact speed $u_0$, maximum force acting on the impactor $F_{\rm max}$, and elapsed time $t_{\rm max}$ to reach $F_{\rm max}$.
We find that  $t_{\rm max}$ emerges in the early stage of the impact, while the rebound process takes place in the late stage.
We find a crossover of $F_{\rm max}$ from $u_0$ independent regime for low $u_0$ to a power law regime satisfying $F_{\rm max}\propto u_0^\alpha$ with $\alpha\approx 1.5$ for high $u_0$.
Similarly, $t_{\rm max}$ satisfies $t_{\rm max}\propto u_0^{\beta}$ with $\beta\approx -0.5$ for high $u_0$.
Both power-law relations for $F_{\rm max}$ and $t_{\rm max}$ versus $u_0$ for high $u_0$ are independent of the system size, but the rebound phenomenon strongly depends on the depth of the container for suspensions.
Thus, we indicate that the rebound phenomenon is not directly related to the relations among $u_0$, $F_{\rm max}$ and $t_{\rm max}$. 
We propose a floating + force chain model, where the rebound process is caused by an elastic term that is proportional to the number of the connected force chains from the impactor to the bottom plate.
On the other hand, there are no elastic contributions in the relations for $F_{\rm max}$ and $t_{\rm max}$ against $u_0$ because of the absence of percolated force chains in the early stage.
This phenomenology predicts $F_{\rm max}\propto u_0^{3/2}$ and $t_{\rm max}\propto u_0^{-1/2}$ for high $u_0$ and also recovers the behavior of the impactor quantitatively even if there is the rebound process.
\end{abstract} 

\maketitle

\section{Introduction}
A dense suspension can behave as a liquid or a solid depending on the situation.
One of the most interesting behaviors of dense suspension is the impact-induced hardening in which the suspension is solidified if the speed of an impactor hitting on a suspension is high enough \cite{brown2014}.
An example of this non-Newtonian behaviors is a running person on the top of a cornstarch suspension, while a walking person sinks~\cite{brown2014}.
This process is practically important for various industrial applications such as protective vests~\cite{lee2003}. 
A similar hardening process is also observable in fractures on a thin layer of a suspension under an impact \cite{roche2013}.
Moreover, this kind of non-equilibrium solid-liquid phase transition is interesting even for physicists.
The impact-induced hardening is often regarded as a process related to the discontinuous shear thickening (DST) under simple shear \cite{lee2003,allen2018}, which attracts much interest among many researchers recently \cite{seto2013,mari2014,townsend2017,sivadasan2019, delgado2020, jamali2020}.
Nevertheless, the underlying mechanism of the impact-induced hardening differs from that of DST as indicated by Ref. \cite{pradipto2020b}.
Indeed, the former is only dominated by the normal stress, while both the normal and shear stresses play important roles in the latter case. 
Thus, impact-induced hardening deserves to be studied on its own.

Let us review some previous studies on the impact-induced hardening.
Waitukaitis and Jaeger conducted an experiment with a rod impactor and discovered the existence of a dynamically jammed region which is a solid plug beneath the impactor~\cite{waitukaitis2012}.
They proposed the added-mass model to explain the solidification induced by the impact.
Then, a series of experiments found that such solidifications take place when the dynamically jammed region is spanned between the impactor and boundaries~\cite{allen2018,maharjan2018,mukhopadhyay2018}. 
The impact-induced hardening can be also observed by dropping an impactor into a dense suspension~\cite{egawa2019, pradipto2020b}.
As a result of the hardening, the free-falling impactor can rebound \cite{egawa2019}.
It is obvious that the elastic effect of dense suspensions is responsible for this rebound phenomenon.

Recently, some papers have discussed the relation between the impact speed $u_0$ and the maximum force acting on the impactor $F_{\rm max}$ or the elapsed time $t_{\rm max}$ to reach $F_{\rm max}$ in impact processes. 
Previous studies\cite{waitukaitis2012,mukhopadhyay2018,brassard2020} showed the existence of power-law relations such as $F_{\rm max}\propto u_0^\alpha$ and $t_{\rm max}\propto u_0^\beta$.
It is noteworthy that similar relations are also found in impact processes for dry granular materials~\cite{krizou2020}.
The numerical solution of the added-mass model~\cite{mukhopadhyay2018} suggests $\alpha=2$ and $\beta=-1$, though the fitted values in their experiment are $\alpha=1.5$ and $\beta=-1/2$.
Moreover, a closer look at the data in Ref. \cite{waitukaitis2012} suggested that $u_0$-independent exponents $\alpha$ and $\beta$ are not appropriate to fit the data in all ranges of the impact speed.
A recent experiment~\cite{brassard2020} also suggested $\alpha=1.5$ and $\beta=-1/2$.
These values of the exponents are obtained as the solution of the viscous force model \cite{brassard2020}, which is inspired by the existence of a growing dynamically jammed region below the impactor \cite{waitukaitis2012, han2015}.
Nevertheless, the viscous force model~\cite{brassard2020} has two defects in which (i) the model cannot explain the behavior for low $u_0$ regime observed in Ref. \cite{waitukaitis2012}, and (ii) the model cannot explain the mechanism of the rebound process since any elastic term is absent.
Therefore, the connection between the rebound of the impactor and the relationships among $u_0$, $F_{\rm max}$, and $t_{\rm max}$ should be clarified to understand the viscoelastic response of an impactor on dense suspensions.

In this paper, we try to clarify the connection between the relations among $u_0$, $F_{\rm max}$, and $t_{\rm max}$ and the rebound phenomena by performing simulations of a free-falling impactor onto dense suspensions based on a coupled model of the lattice Boltzmann method (LBM) and discrete element method (DEM).
In addition, we propose a phenomenology to explain these processes, including the elastic force as a result of percolated force chains between the impactor and bottom plate to describe the rebound phenomenon.
This model is essentially reduced to the viscous force model if percolated force chains are absent.

The outline of the paper is as follows.
In Sec. \ref{sec:method}, we briefly explain the method and setup of our simulation.
In Sec. \ref{sec:results}, we present the results of our simulation including the impactor motion and the relationships among $u_0$, $F_{\rm max}$ and $t_{\rm max}$.
In Sec. \ref{sec:model}  we examine our phenomenology with and without the elastic force between the impactor and bottom plate to explain the results of the simulation.
In Sec. \ref{sec:conclusion}, we summarize our results and discuss future perspectives.
In Appendix \ref{app:lbm}, we describe the details of our simulation method.
In Appendix \ref{app:sol}, we present the exact solution and its approximate treatment of our phenomenology when the elastic force is absent.
In Appendix \ref{app:fc}, we describe the details of the force chains analysis in the phenomenology.
Finally in Appendix \ref{app:dilute}, we discuss the dependence of our results on the volume fraction of the suspensions.

\section{\label{sec:method} Setup of our simulation}

\begin{figure}[htbp]	
	\includegraphics[width=0.7\linewidth]{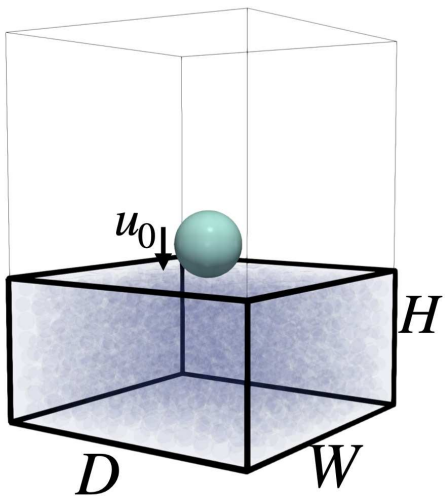}
	\caption{An illustration of an initial setup of our simulation.}
	\label{fig:ilus1}
\end{figure}

\begin{figure*}[htbp]
	\includegraphics[width=\linewidth]{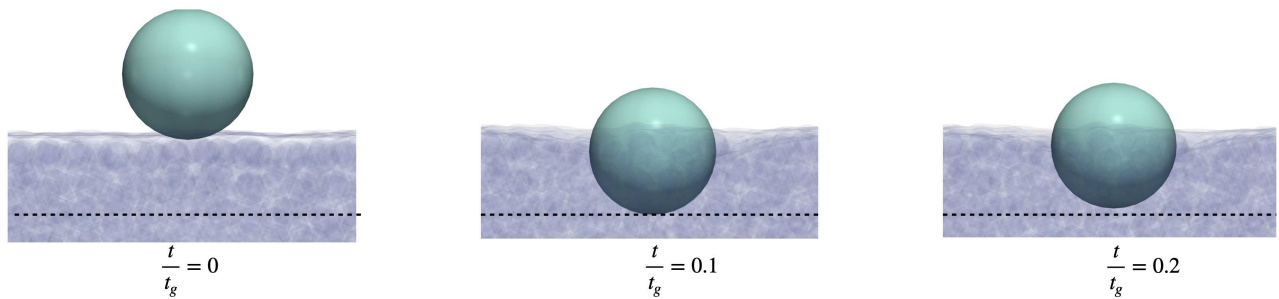}
	\caption{Successive snapshots of an impactor in a quasi-two-dimensional projection, where the black dashed lines correspond to the maximum penetration (deepest position) of the impactor.}
	\label{fig:ilus3}
\end{figure*}
\begin{table}[htbp]
	\centering
	\begin{tabular}{ c | c | c | c} 
		\hline	\hline
		Volume fraction $\phi$ & Depth $H$ & Width $W$ & No. of particles $N$ \\ 
		\hline
		0.00	& $3D_I$ & $6D_I$ &  0\\ 
		\hline
		0.10	& $3D_I$ & $6D_I$ &  409 \\ 
		\hline
		0.25	& $3D_I$ & $6D_I$ &  1021 \\ 
		\hline	
		0.40 	& $3D_I$ & $6D_I$ &  1634 \\ 
		\hline
		0.48 	& $3D_I$ & $6D_I$ &  1960 \\ 
		\hline	
		& $2D_I$ & $4D_I$ &  617  \\ 
		0.51 	& $3D_I$ & $6D_I$ &  2083 \\ 
		& $6D_I$ & $5D_I$ &  2893 \\ 
		& $7D_I$ & $4D_I$ &  2160 \\ 
		\hline
		& $2D_I$ & $4D_I$ & 642 \\ 
		0.53	& $2D_I$ & $6D_I$ & 1443 \\ 
		& $3D_I$ & $6D_I$ & 2164 \\ 
		& $7D_I$ & $4D_I$ & 2245  \\ 
		\hline
		0.56 & $2D_I$ & $4D_I$ & 677  \\
		& $7D_I$ & $4D_I$ & 2371 \\  		
		\hline	\hline
	\end{tabular}
	\caption{All variations of simulated volume fractions $\phi$ and box sizes with the corresponding numbers of suspended particles $N$.}
	\label{tab:param}	
\end{table}

We simulate a binary mixture of suspension consisting of equal number of large and small particles with bidispersity ratio $a_{\text{max}} = 1.2 a_{\text{min}}$, where the radii of the large and small particles are $a_{\text{max}}$ and $a_{\text{min}}$, respectively. 
We use the mixture to avoid the crystallization of suspended particles in high density regions.
These suspended particles have the identical density $\rho_p$ which is equal to the density $\rho_f$ of the solvent.
The suspension liquid is confined in a rectangular box surrounded by smooth sidewalls and a smooth bottom plate. 
Since we simulate free falling processes of an impactor, there is no lid above the container.   
The volume $V$ of the suspension liquid at rest is expressed as $V = W \times D \times H$, where $H$ is the depth of the suspension and $W=D$ is the width of the container as shown in Fig. \ref{fig:ilus1}.
The volume fraction $\phi$ of the suspension at rest without the impactor is defined as $\phi = 2N \pi (a_{\text{max}}^{3} + a_{\text{min}}^{3})/ 3V$, where $N$ is the number of suspended particles used in the simulation.
The hydrodynamic interaction among particles is simulated using the LBM.
The contact force between suspended particles is modeled by the DEM \cite{luding2008} with the spring constant $k_n$ between contacting particles.
We also introduce the frictional contact model between suspended particles with Coulomb's friction rule, which is important to recover the hardening behavior of dense suspensions \cite{mari2014,townsend2017,pradipto2020,pradipto2020b}.
In this paper, we adopt the friction coefficient $\mu=1$ for all cases.
As it is known~\cite{sivadasan2019,pradipto2020b}, the rheological properties of dense suspensions is insensitive to $\mu$ for $\mu \geq 0.3$ .
Details of our simulation method can be seen in Appendix \ref{app:lbm}.

A spherical impactor with diameter $D_I$ and density $\rho_I$, is released from the height $H_0$ which corresponds to the impact speed $u_{0} = \sqrt{2 g H_0}$ with the gravitational acceleration $g$.  
In our simulation $\rho_I$ and $D_I$ satisfy $\rho_I= 4 \rho_f$ and $D_I=6 a_{\rm min}$, respectively.
We also introduce the time scale $t_g = \sqrt{a_{\text{min}}/2g}$, speed scale  $u^{*} = \sqrt{2 g a_{\text{min}}}$, and force scale $F_g=\frac{4}{3} \pi \rho_f (D_{I}/2)^3 g$.
It should be noted that there is another important time scale $t_k=\sqrt{m_0/k_n}$ where $m_0=\frac{4}{3}\pi \rho_f a_{\rm min}^3$.
Thus, the mass of the impactor $m_I$ is expressed as $m_I=\frac{\pi}{6}\rho_I D_I^3$.
In our simulation the ratio $t_k/t_g=0.045$ is fixed. 
Note that we also adopt the DEM for the contact interactions between the impactor and suspended particles, between suspended particles and container's walls, and between the impactor and container's walls.
All variations of simulated volume fractions $\phi$ and box sizes are summarized in Table. \ref{tab:param}. 
Note that we use three ensembles for $\phi=0.53$, $W=D=6 D_I$, and $H=3 D_I$ and for $\phi=0.53$, $W=D=6 D_I$, and $H=2 D_I$.
We only simulate one ensemble for the other cases.

\section{\label{sec:results} Simulation results}

\subsection{\label{sub:imp} Impactor motion}

\begin{figure}[htbp]
	\includegraphics[width=0.95\linewidth]{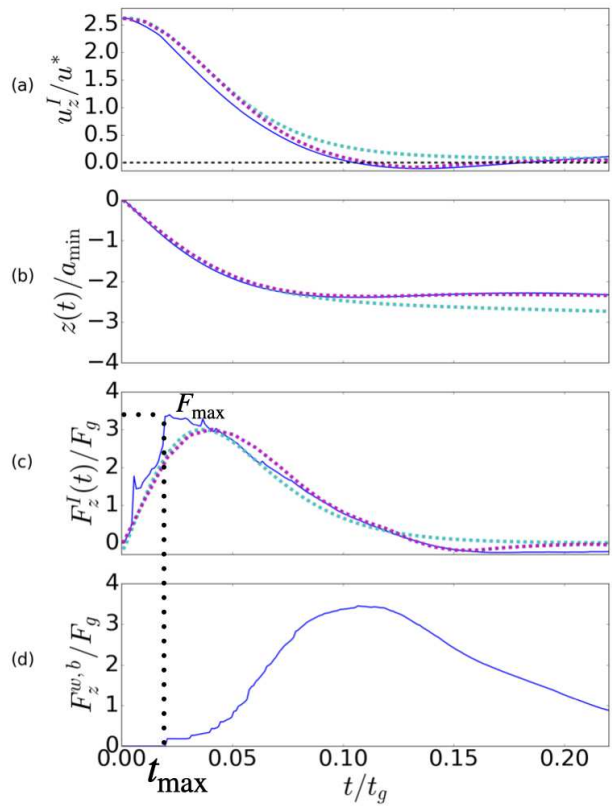}
	\caption{Plots of the time evolutions of the impactor motion (blue solid lines) for $\phi=0.53,$ $ W=D=6 D_I$, and $H=2D_I$ for (a) the velocity $u^{I}_{z}/u^{*}$ (black dashed line represents $u^{I}_{z}/u^{*} = 0$), (b) the position of the deepest point of the impactor $z(t)/a_{\rm min}$, (c) the force exerted on the impactor $F^{I}_{z}/F_g$, and (d) the force exerted on the bottom plate $F^{w,b}_{z}/F_g$, respectively. Dashed purple lines in (a), (b), and (c) represent the solution of Eq. \eqref{eq:eom_fc} and dashed light blue lines in (a), (b), and (c) represent the solution of Eq. \eqref{eq:eom}. Black dotted lines highlight $F_{\rm max}$ and $t_{\rm max}$.}
		\label{fig:1a}
			\label{fig:1b}
				\label{fig:1c}
					\label{fig:1d}
	\label{fig:1}
\end{figure}

\begin{figure}[htbp]
	\includegraphics[width=0.8\linewidth]{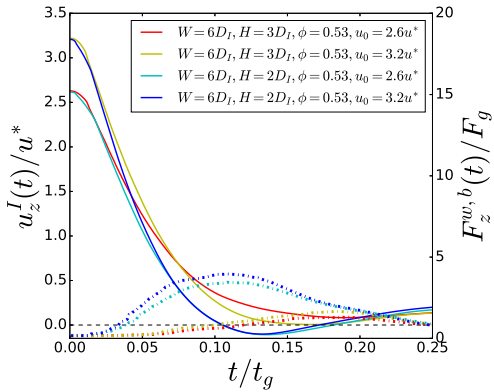}
	\caption{Plots of the time evolutions of the impactor velocities $u^{I}_{z}/u^{*}$ for $\phi=0.53$ and $ W=D=6 D_I$ (left vertical axis) for various $u_0$ and $H$.
			Blue and cyan solid lines represent the results for $H=2 D_I$ with $u_0/u^{*}=3.2$, and $H=2 D_I$ with $u_0/u^{*}=2.6$, respectively. 
			Red and yellow solid lines represent the results for $H=3 D_I$ with $u_0/u^{*}=3.2$, and $H=2 D_I$ with $u_0/u^{*}=2.6$, respectively. 
			Black dashed line represents $u^{I}_{z}/u^{*} = 0$. 
			The dot-dashed lines represent the corresponding forces exerted on the bottom plate $F^{w,b}_{z}/F_g$ (right vertical axis).}
	\label{fig:deeplow}
\end{figure}

\begin{figure*}[htbp]
	\subfloat[]{\label{fig:2a}%
		\includegraphics[width=0.4\linewidth]{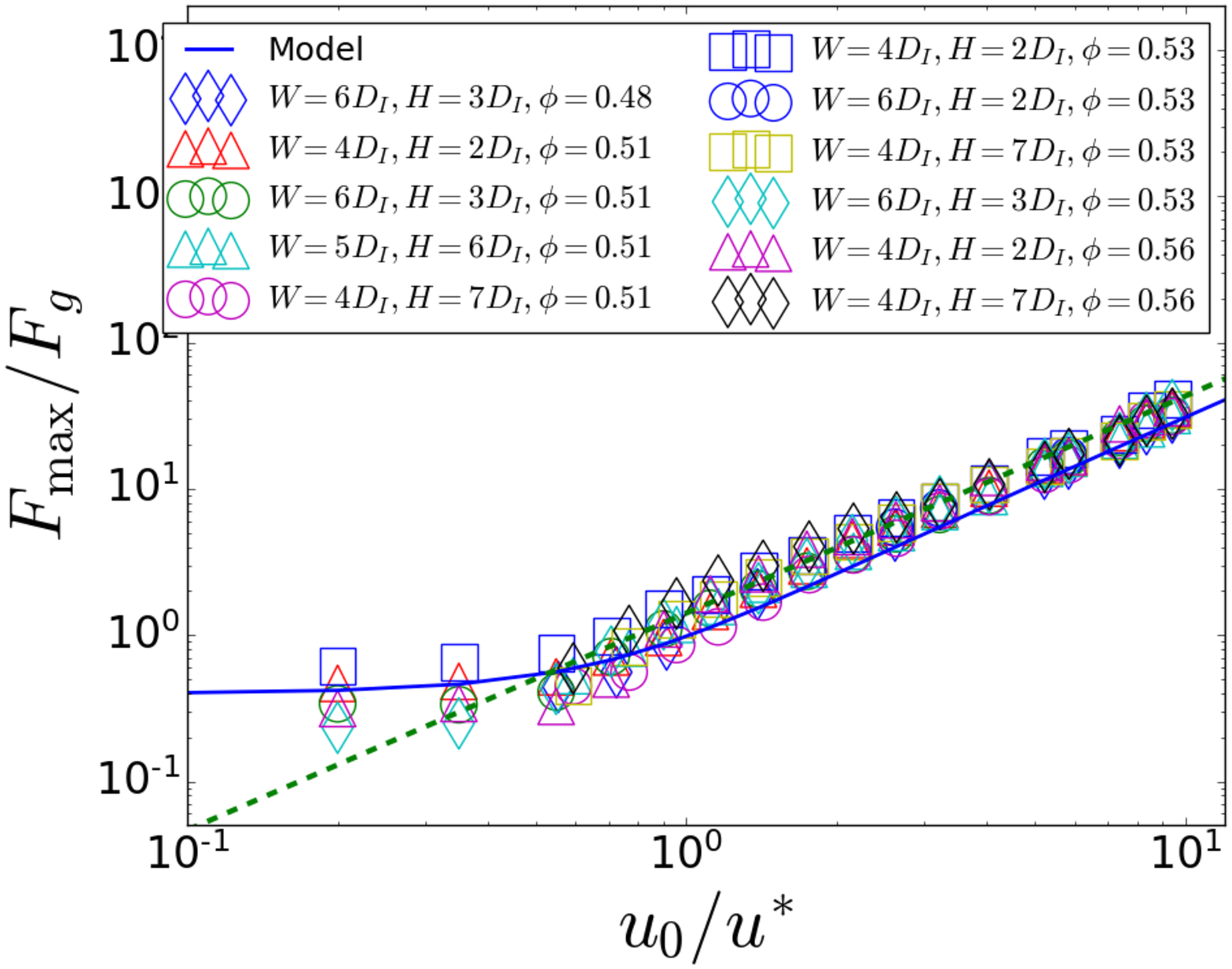}
	}
	\subfloat[]{\label{fig:2b}%
		\includegraphics[width=0.4\linewidth]{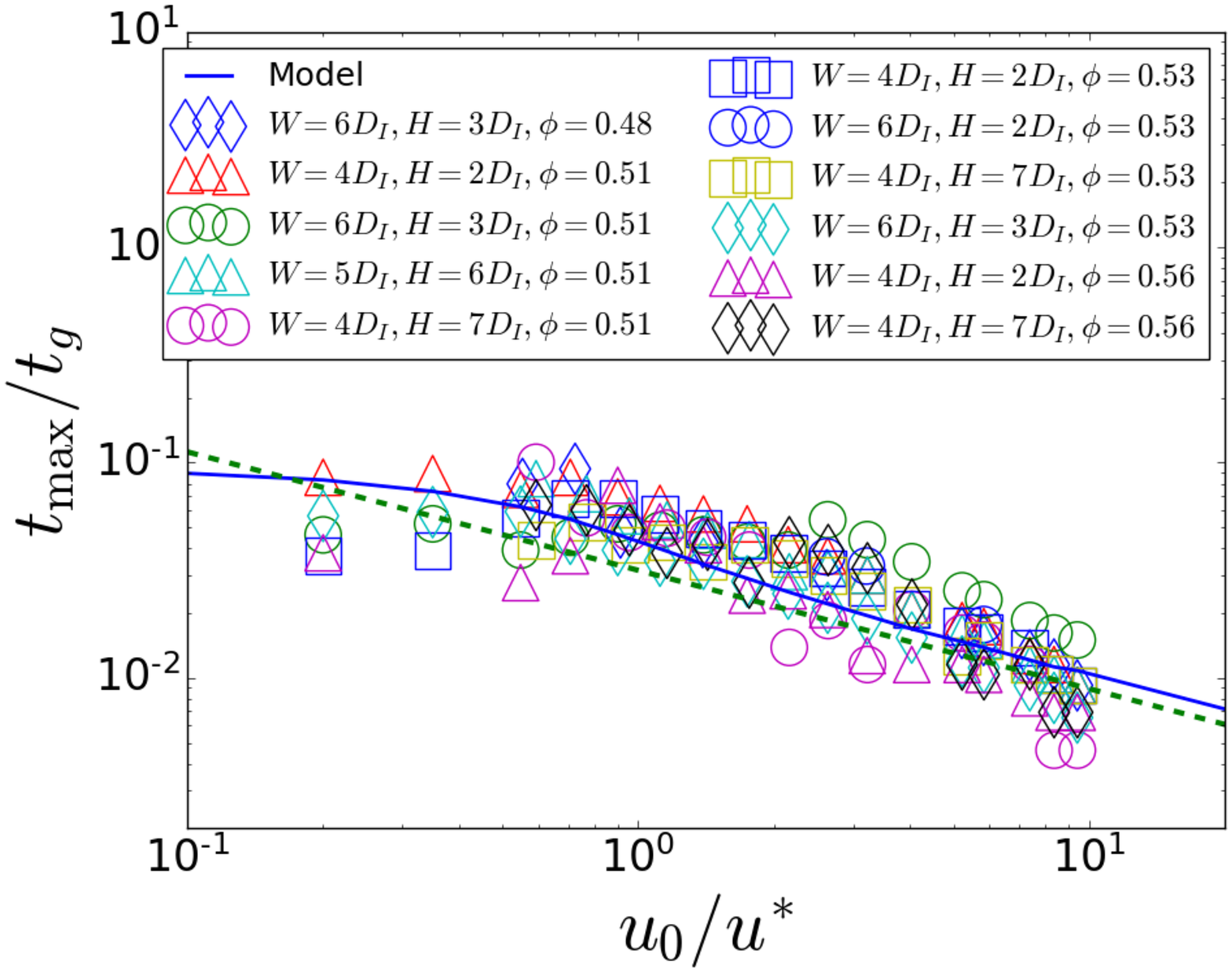}
	}
	\caption{(a) Plots of maximum forces exerted on the impactor $F_{ \rm max}$ scaled by the gravitational force $F_g$ against $u_0/u^{*}$ for various container sizes, where the green dashed line represents  $1.58(u_0/u^*)^{1.432}$. (b) Plots of time $t_{\rm max}$ to reach $F_{ \rm max}$ scaled by $t_g$ against $u_0/u^{*} $, where the green dashed line represents  $0.03(u_0/u^*)^{-0.523}$. The blue solid lines in both figures represent the solution of the floating model (Eq. \eqref{eq:eom}).}
	\label{fig:2}
\end{figure*}

Figure \ref{fig:ilus3} shows successive snapshots of an impactor in a quasi-two-dimensional projection of our three-dimensional simulation, where the black dashed lines correspond to the maximum penetration position of the impactor.
We can also identify a rebound process from the middle and right figures in which the vertical position of the impactor in the right figure is higher than that in the middle.

In Fig. \ref{fig:1}, we plot the time evolutions of the velocity and vertical position of the impactor, force acting on the impactor, and force acting on the bottom plate obtained from our simulation.
From Fig.3(a), one can see the existence of a rebound process i.e.\ the region for $u_z^I < 0$.
As a result, the vertical position $z(t)$ increases with time in the rebound process as shown in Fig. 3(b).
From Fig. 3(c), one can define $F_{\rm max}$ as its peak value and $t_{\rm max}$ as the time to reach $F_{\rm max}$.
Note that $t_{\rm max}$ coincides with the onset time of the force exerted on the bottom plate $F^{w,b}_z$, while the rebound takes place around and after the peak of $F^{w,b}_z$ (see Fig. 3(d)). 
This indicates that the rebound of the impactor takes place when the force from the impactor is transmitted through the force chains to the bottom plate.
In Sec. \ref{chain}, we will confirm this picture.

We also find that $t_{\rm max}$ is located much earlier than time of the rebound region ($u_z^I < 0$).
This suggests that $F_{\rm max}$ and $t_{\rm max}$ are not related to the rebound motion.
This is consistent with the following observation in which $F_{\rm max}$ and $t_{\rm max}$ are independent of system size \cite{brassard2020} but the rebound motion strongly depends on the system size.
Indeed, as can be seen in Fig. \ref{fig:deeplow}, the rebound takes place only for the suspension in a shallow vessel as in the case of $H=2D_I$, while the rebound cannot be observed for the suspension in a deep vessel ($H=3D_I$).
At the early stage for $t/t_g <0.05$, such depth dependence does not exist.

\subsection{\label{sub:scl}  Relations among $u_0$, $F_{\rm max}$ and $t_{\rm max}$}

In Fig. \ref{fig:2}, we plot $F_{\rm max}$ exerted on the impactor scaled by the gravitational force $F_g$ against $u_0$ for $\phi \geq 0.48$ (see Appendix \ref{app:dilute} for the results of $\phi \leq 0.40$).
Here, the results of our simulation for $F_{\rm max}$ and $t_{\rm max}$ show the existence of power-law regimes satisfying
\begin{equation}
	F_{\rm max}\propto u_0^\alpha, \quad t_{\rm max}\propto u_0^\beta
	\label{eq:scl}
\end{equation}
with $\alpha = 1.432 \pm 0.0003$ and $\beta = -0.523  \pm 0.042$ for $u_0>u^{*}$.
One can find that the data for all volume fractions and system sizes are collapsed on a universal curve for $F_{\rm max}$, while $t_{\rm max}$ does not have the beautiful data collapse.
Our observed exponents agree with those in the experiment \cite{waitukaitis2012,brassard2020} and is smaller than the solution of the added-mass model ~\cite{mukhopadhyay2018}.
The values of $\alpha$ and $\beta$ also are close to those obtained by the viscous force model \cite{brassard2020}. 
This is understandable since the peak of the force exists in the early stage where the elastic force to produce the rebound does not play any role.
Thus, one does not need to take into account the elastic force to explain the relations among $u_0$, $F_{\rm max}$, and $t_{\rm max}$.
Moreover, we have simulated variations of widths and depths in Fig. \ref{fig:2} to confirm that the relations among $u_0$, $F_{\rm max}$, and $t_{\rm max}$ are independent of the system size.
This is in contrast to the rebound phenomenon which strongly depends on the width and depth of the simulation box (see Fig. \ref{fig:deeplow}).
This observation is another evidence that the relations among $u_0$, $F_{\rm max}$, and $t_{\rm max}$ are not related to the rebound phenomenon.
 
Our simulation also illustrates that a single power-law is insufficient for $F_{\rm max}$ versus $u_0$ to fit the data in all ranges of the impact speed.
Instead, we find a crossover of the relation between $F_{\rm max}$ and $u_0$ from $u_0$ independent regime for low $u_0$ to the power-law region for high $u_0$ regime (see Fig. \ref{fig:2a} ).
The corresponding $u_0$ independent regime of $t_{\rm max}$ for low $u_0$ is also visible in Fig. \ref{fig:2b}, though the data are not clear enough. 
Even though the authors of Ref. \cite{waitukaitis2012} did not mention such a crossover in their paper, their data suggest the existence of a subtle crossover in the relation between $u_0$ and $F_{\rm max}$, similar to what we have observed.
Furthermore, it is obvious that a set of single values of $\alpha$ and $\beta$ is no longer valid if the acceleration due to gravity plays some roles, as will be shown in the next section.
This might be the reason why the viscous force model in Ref. \cite{brassard2020} cannot explain the existence of $u_0$ independent regime.

\section{\label{sec:model} PHENOMENOLOGY}

\subsection{Overview}

Judging from the observations in our simulation, we propose the following simple phenomenology to describe the vertical motion of an impactor:
\begin{equation}
		m_I \frac{d^2z_I}{dt^2}= -m_I \tilde{g}  +  F^{I}_{D},
		\label{eq:2}
	\end{equation}
where $z_I(t)$ is the vertical position of the center of mass of the impactor, $\tilde{g}$ is the effective gravitational acceleration defined as $\tilde{g} = g(\rho_I - \rho_f)/\rho_f$, and  $F^{I}_{D}$ is the drag force acting on the impactor.
It should be noted that Ref. \cite{pradipto2020b} adopted the dynamical Hertzian contact model (DHCM) but the predictions of the DHCM, $6/5 < \alpha < 4/3$ and $-1/3 < \beta < -1/5$, disagree with the simulation and experimental results.
DHCM has also another drawback in which it cannot recover the $u_0$ independent regime observed in our simulation.

Recently, Brassard et al. \cite{brassard2020} proposed the viscous force model including a drag term which is proportional to the depth of the impactor, though their model ignores the gravity term $m_I \tilde{g}$ and the elastic force to reproduce the rebound process.
Although their model cannot explain $u_0$ independent regime and the rebound process, the analytic solution of the model yields $\alpha=1.5$ and $\beta=-0.5$.
Our proposed model in the early stage is essentially the same as that in Ref. \cite{brassard2020} with keeping the gravity term (Sec. \ref{flo}).
Of course, we should take into account the elastic force in the late stage if there are percolated force chains from the impactor to the bottom plate (Sec. \ref{chain}).
The model used in Sec.  \ref{chain} reduces to the model in Sec. \ref{flo} because the former contains the number of percolated force chains $n(t)$ which becomes zero in the latter case.
Nevertheless, we will explain the floating model with $n(t)=0$ in the early stage in Sec. \ref{flo}, and introduce the floating + force chain model with $n(t)\ne 0$ in the late stage in Sec. \ref{chain} separately. 

\subsection{Floating model \label{flo}}

\begin{figure}[htbp] 
	\includegraphics[width=0.75\linewidth]{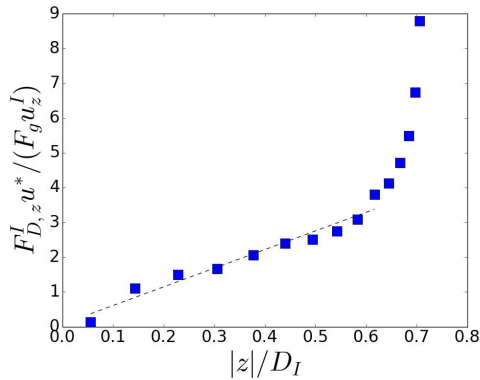}
	\caption{Plot of the drag exerted on the impactor $F^{I}_{D,z}$ scaled by $F_g$ and impactor velocity $u^{I}_{z}/u^{*}$ against impactor depth scaled by diameter of the impactor $z_I/D_I$. Black dashed line represents the linear fit of $F^{I}_{D,z}/F_g$ for  $z_I/D_I \leq 0.7$ }
	\label{fig3}
\end{figure}

Let us propose a simple phenomenology which we call the floating model to explain the behavior of the impactor for both $F_{\rm max}$ and $t_{\rm max}$ in the early state.
To model the motion of the impactor, we assume that the impactor is only influenced by the gravity and viscous drag force from the surrounding suspension in the early stage.
This assumption is based on the observation that the dynamically jammed region is floating without touching the bottom plate in the early stage of the impact \cite{waitukaitis2012,han2015}. 
We also assume that the drag force is proportional to the impactor velocity because the fluid drag should be determined by the Stokes flow.
Thus, in order to extract the coefficient, we plot the drag exerted on the impactor $F_D^I$  divided by the velocity against $|z|$ in Fig. \ref{fig3}, where $z$ is the deepest position of the impactor ($z=0$ is the instance of attachment of the impactor on the surface of the liquid). 
Here, we confirm that the drag is proportional to the impactor depth when $|z|/D_I \leq 0.7$.
If we assume that $F_D^I$ is proportional to $|z| \dot{z}_I$, $F_D^I$ in Eq. \eqref{eq:2} is identical to the drag force in the viscous force model \cite{brassard2020}.
It should be noted that the center of mass of the impactor $z_I$ is related to $z$ as $z_I=z+a_I$.
Then, the total drag force acting on the impactor is given by
\begin{equation}
F^{I}_{D} =3\pi \eta_{\rm eff}  \dot{z}_I |z|,
\label{eq:drag}
\end{equation}
where we have introduced the effective viscosity $\eta_{\rm eff}$ to characterize the apparent viscosity of the dynamically jammed region.
The derivation of Eq. \eqref{eq:drag} can be seen in Appendix \ref{app:sol}.
Then, we can write the equation of motion of the impactor as
\begin{equation}
	m_I \frac{d^2z_I}{dt^2} = -m_I \tilde{g} +  3\pi \eta_{\rm eff}  \dot{z}_I |z|.
	\label{eq:eom}
\end{equation}
Equation \eqref{eq:eom} can be solved exactly with the aid of the Airy functions (see Appendix \ref{app:sol}). 
The solutions for $z_I(t)$, $u_z^I(t)=-dz_I(t)/dt$, and $F_z^I(t)=m_I d^2z_I(t)/dt^2$ are plotted alongside the simulation results in Fig. \ref{fig:1}.
The numerical solutions for $F_{\rm max}$ and $t_{\rm max}$ (the blue solid lines) are presented in Figs. \ref{fig:2a} and  \ref{fig:2b}, respectively.
We use the value of the effective viscosity $\eta_{\rm eff}=4.9 \times 10^{4} m_0/(a_{\rm min} t_g)$ as a fitting parameter.
This value is about a hundred times larger than the viscosity of the solvent $\eta_0$ and five times larger than the observed viscosity for DST under simple shear using LBM simulation \cite{pradipto2020} (see Appendix \ref{app:dilute}).
The enhancement of the viscosity is reasonable, because the impactor contacts with the dynamically jammed region which must have larger viscosity than that of the averaged suspension.
One can see that  Eq. \eqref{eq:eom} can recover the crossover from $u_0$ independent regime for low $u_0$ to the power law regime for high $u_0$ observed in our simulations.
The solution of Eq. \eqref{eq:eom} yields 
\begin{equation}
\alpha = \frac{3}{2}, \quad \beta = -\frac{1}{2}
\label{eq:exp}
\end{equation}
for high $u_0$ (see Appendix B for how to obtain these exponents).
$F_{\rm max}$ should be independent of $u_0$ for low $u_0$ because the second term on the right-hand side (r\@.h\@.s\@.) of Eq. \eqref{eq:eom} is much smaller than the first term for low $u_0$.
This is the simple explanation for the crossover observed in Fig. \ref{fig:2}.

\subsection{Floating + force chains model \label{chain}}

\begin{figure}[htbp] 
	\includegraphics[width=0.75\linewidth]{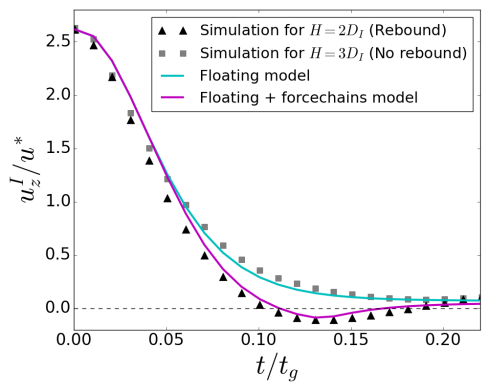}
	\caption{Plots of time evolutions of velocities of the impactors for a rebound case with $H=2D_I$ and a no-rebound case with $H=3D_I$. 
		Filled triangles represent the simulation results of $\phi=0.53,$ $ W=D=6 D_I$, and $H=2 D_I$. 
		The green solid line represents the solution of Eq. \eqref{eq:eom}, and the purple solid line represents the solution of Eq. \eqref{eq:eom_fc} (the black dashed line represents $u^{I}_{z}/u^{*} = 0$). Here, we also plot the simulation results for $\phi=0.53$, $W=D=6 D_I$, and $H=3 D_I$ (black squares), where  rebound does not take place.}
	\label{fig5}
\end{figure}

Unfortunately, Eq. \eqref{eq:eom} cannot explain the rebound of the impactor because of the absence of elastic force which is the origin of the rebound (see Fig. \ref{fig:1a}).
This indicates the drawback of the viscous force model which cannot explain the rebound process.
We also note that the solution of  Eq. \eqref{eq:eom} is independent of system size, which is consistent with the results of Ref. \cite{brassard2020} and Fig. \ref{fig:2} in this paper.
On the other hand, our simulation in Sec. \ref{sub:imp} indicates that rebound is related to the force acting on the bottom plate.
Since the force can be transmitted via contacts of suspended particles along the chains, we can calculate the elastic force along the chains (see Appendix \ref{app:fc} for definition and visualizations of the force chains).
Then, we include an elastic term to Eq. \eqref{eq:eom} caused by connected force chains between the impactor and bottom plate as 
\begin{equation}
	m_I \frac{d^2z_I}{dt^2}=  -m_I \tilde{g} +  3\pi \eta_{\rm eff}  \dot{z}_I |z| + n(t) k_n z_I,
	\label{eq:eom_fc}
\end{equation}
where $n(t)$ is the number of connected chains  from the impactor to the bottom plate, and $k_n$ is the spring constant of the DEM.
In other words, the elastic force (the third term on the r\@.h\@.s\@. of Eq. \eqref{eq:eom_fc}) is originated from the contacting elastic force along the force chains of contacting suspended particles between the impactor and bottom plate.
Details of the algorithm to determine $n(t)$ is written in Appendix \ref{app:fc} and is illustrated in the Fig. \ref{fig:fc} (Multimedia view).
It is obvious that Eq. \eqref{eq:eom_fc} is reduced to Eq. \eqref{eq:eom} if the percolated force chains do not exist, i. e. $n(t)=0$ in the early stage.
In this sense, the model in Eq.  \eqref{eq:eom_fc} is more general than the floating model described by Eq. \eqref{eq:eom}. 

In Fig. \ref{fig5}, we plot time evolutions of the impactor velocity from our simulation alongside with the corresponding results of Eqs. \eqref{eq:eom} and \eqref{eq:eom_fc} with  $\eta_{\rm eff}=4.9 \times 10^{4} m_0/(a_{\rm min} t_g)$ and $k_n = 2.5 \times 10^{4}  m_0/(a_{\rm min} t_{g}^{2})$ which is identical to that used in the DEM simulation.
Here, one can see that the rebound of the impactor can be recovered by the introduction of the third term on the r.h.s.\ of Eq. \eqref{eq:eom_fc} for the shallow vessel case ($H=2D_I$) (see Fig. \ref{fig5}).
On the other hand, the floating model (Eq. \eqref{eq:eom}) is sufficient to recover the impactor velocity correctly for the deep vessel case ($H=3D_I$) where rebound does not take place.
Thus, the phenomenology described by Eq. \eqref{eq:eom_fc} can describe the quantitative behavior of the impactor by the introduction of two fitting parameters $\eta_{\rm eff}$ and $n(t)$, though $n(t)$ is determined by the observation as shown in Appendix \ref{app:fc}.
Thus, our phenomenology is more accurate than the linear model in Ref. \cite{egawa2019} and the DHCM in Ref. \cite{pradipto2020b}.
Note that the power-law exponents $\alpha$ and $\beta$ in \eqref{eq:exp} are not affected by this elastic term since $F_{\rm max}$ and $t_{\rm max}$ emerge in the early stage of the impact.

\section{\label{sec:conclusion} Discussion and Conclusions}

We numerically studied the impact processes on dense suspensions using a coupled model of LBM and DEM to elucidate the connection between the elastic rebound of the impactor and the relations among $u_0$, $F_{\rm max}$, and $t_{\rm max}$.
Then, we have also proposed a simple phenomenology called the floating+force chain model to explain our simulation results.
This model reduces to the floating model if there are no percolated force chains from the impactor to the bottom plate.
We numerically find the existence of a power-law regime satisfying $F_{\rm max}\propto u_0^\alpha$, with $\alpha=1.432 \pm 0.0003$ and $t_{\rm max}\propto u_0^\beta$, with $\beta = -0.523  \pm 0.042$, while the analytic solution of the floating model indicates $\alpha=3/2$ and $\beta=-1/2$.
We have also confirmed the existence of $u_0$-independent regimes of $F_{\rm max}$ and $t_{\rm max}$ for low $u_0$.
The crossovers of $F_{\rm max}$ and $t_{\rm max}$ from $u_0$-independent regimes to the power law regimes can be reproduced by the floating model correctly.
We conclude that the relations among $u_0$, $F_{\rm max}$, and $t_{\rm max}$ are not related to the rebound process based on three observations:
(i) We found that $F_{\rm max}$ emerges in the early stage of the impact, while the rebound of the impactor takes place in the later stage.
(ii) We have confirmed that the relations among $u_0$, $F_{\rm max}$, and $t_{\rm max}$ are independent of the system size, while the rebound strongly depends on the size of the container.
(iii) One can recover the exponents for $F_{\rm max}$ and $t_{\rm max}$ when the impactor depth is smaller than its diameter (not completely sink) in the drag term that is proportional to the impactor depth without considering any elastic force, which agrees with Ref. \cite{brassard2020}.
In contrast, the rebound needs an elastic term caused by the connected force chains from the impactor to the bottom plate.

Our phenomenology, the floating + force chain model, is only valid for dense suspensions because any percolated force chains do not exist for dilute suspensions. 
Nevertheless, $F_{\rm max}$ and $t_{\rm max}$ can exist even for dilute suspensions.
Although the velocity of the impactor can be partially explained by a model in which $F_D^I$ is expressed as the Stokes drag force, the Stokes drag model cannot explain the existence of $F_{\rm max}$ and $t_{\rm max}$ in dilute situations (see Appendix \ref{app:dilute}).
To estimate $\eta_{\rm eff}$ within our phenomenology we need to take into account the interaction between the impactor and the dynamically jammed region.
However, the dynamically jammed region is still not well-defined, and thus, this may not be a well-defined problem.
Furthermore, our algorithm to determine the number of percolating force chains from the impactor to the bottom plate ($n(t)$) essentially ignores the role of the sidewalls.
This needs to be improved if one examines the impact-induced hardening phenomena in a narrow channel or using bumpy sidewalls.
We only focused on relatively short time behavior after the impact, while a sinking impactor in dense suspensions shows a distinct behavior, as it oscillates and exhibits a stop-go cycle near the bottom of the container \cite{vonkann2011}.
Our simulation will be able to be used to reproduce these results. 

\begin{acknowledgments}
One of the authors (P) expresses his gratitude to Alessandro Leonardi for sharing his lattice Boltzmann code.
The authors thank Satoshi Takada for his critical reading of the manuscript.
All numerical calculations were carried out at the Yukawa Institute for Theoretical Physics (YITP) Computer Facilities, Kyoto University, Japan.
This research is partially supported by  Grants-in-Aid of MEXT, Japan for Scientific Research, Grant Nos. JP16H04025 and JP21H01006.
\end{acknowledgments}

\appendix

\section{\label{app:lbm} LBM + DEM with free surface}
 
  \begin{figure*}[htbp] 
 	\includegraphics[width=\linewidth]{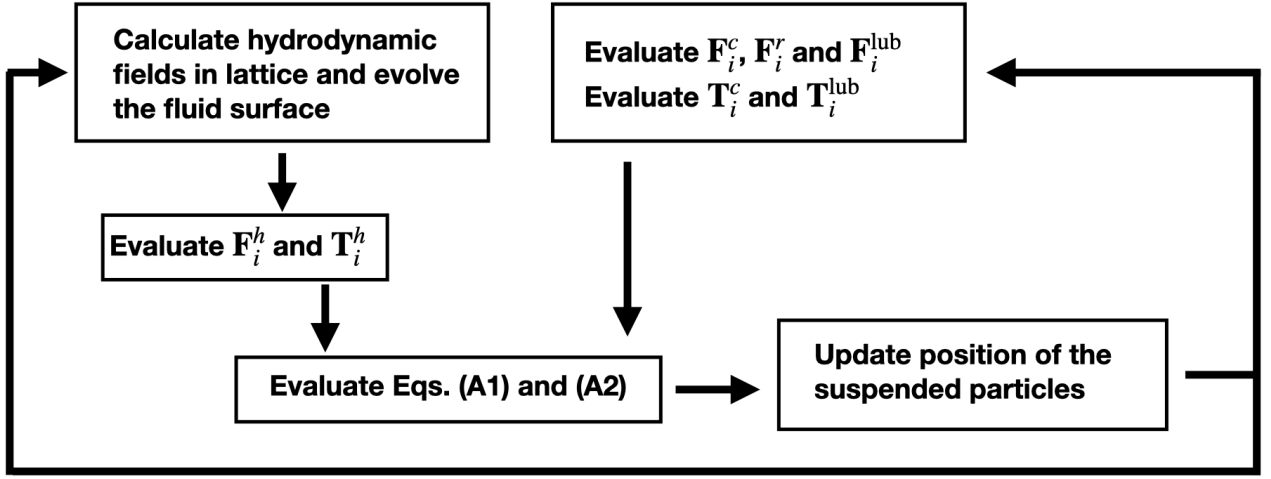}
 	\caption{Summary of the LBM + DEM simulations.} 
 	\label{fig:flowchart}
 \end{figure*}

 \begin{figure}[htbp] 
	\includegraphics[width=0.7\linewidth]{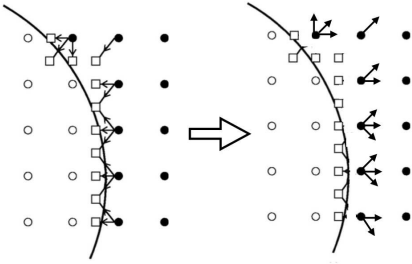}
	\caption{An illustration of the bounce-rule in LBM simulation. Filled circles represent fluid nodes, open circles represent solid nodes, open squares represent boundary nodes, arrows represent the streaming discrete distribution functions.} 
	\label{fig:bb}
\end{figure}

 \begin{figure}[htbp]
	\includegraphics[width=0.7\linewidth]{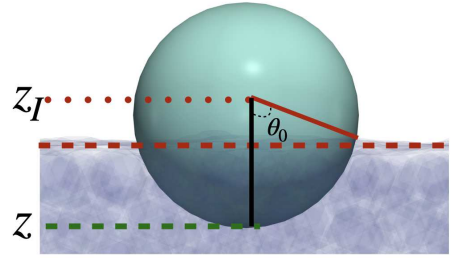}
	\caption{An illustration of an impactor in the suspension liquid to explain $z$, $z_I$, and $\theta_0$   (the black solid line connects the deepest position of the impactor at $z$ with the center of mass at $z_I$, the red solid line is the line between the surface of the suspension and the center of mass, the green dashed line represents the deepest point  $z$, the red dashed line is the surface of the suspension ($z=0$), and $\theta_0$ is the angle between the black and red solid lines. }
	\label{fig:ilus2}
\end{figure}

We employ the LBM involving suspensions and the free surface of the fluid.
The summary of this method can be seen in Fig. \ref{fig:flowchart}.
Throughout this paper, we have adopted the perfect density matching between the solvent and suspended particles, where the densities of particles and solvent satisfy the relation $\rho_p=\rho_f$, where $\rho_p$ and $\rho_f$ are the densities of a suspended particle and solvent fluid, respectively.
The details of the LBM are explained in Ref.~\cite{pradipto2020b}.
The suspended particles in LBM are represented as a group of solid nodes, while the surrounding fluids are represented by fluid nodes.
The hydrodynamic field is calculated from the time evolution of the discrete distribution function at each fluid node.
We select the lattice unit $\Delta x = 0.2 a_{\text{min}}$, where it gives sufficient accuracy but still not computationally expensive as shown in the previous LBM for suspensions literatures \cite{ladd1994a,ladd1994b,nguyen2002}.
In addition, to simulate the free surface of the fluid, it is necessary to introduce interface nodes between the fluid and gas nodes \cite{svec2012,leonardi2014,leonardi2015, pradipto2020b}.

Equations of motion and the torque balance of particle $i$ are, respectively, given by 
\begin{equation}
	m_i \frac{d \bm{u}_i}{dt} = \bm{F}_i^c + \bm{F}_i^h + \bm{F}_i^{\rm lub} + \bm{F}_i^r + \bm{F}_i^g,
	\label{eom_part}
\end{equation}
\begin{equation}
	I_i \frac{d \bm{\omega}_i}{dt} = \bm{T}_i^c +  \bm{T}_i^{\rm lub} + \bm{T}_i^{h}.
	\label{eot_part}
\end{equation}
Here, $\bm{u}_i$, $\bm{\omega}_i$, $m_i$, and $I_i=(2/5) m_i a_i^2$ (with $a_i$ the radius of particle $i$), are the translational velocity, angular velocity, mass, and the moment of inertia of particle $i$, respectively.
$\bm{F}_i^g = - m_i g \hat{\bm{z}}$ is the gravitational force acting on the suspended particle $i$, where $\hat{\bm{z}}$ is the unit vector in the vertical direction. 

Note that our LBM accounts for both the short-range lubrication force $\bm{F}_i^{\rm lub}$ and torque $\bm{T}_i^{\rm lub}$, as well as the long-range hydrodynamic force $\bm{F}_i^h$ and torque $\bm{T}_i^h$ as in Ref.~\cite{nguyen2002,pradipto2020}.
The long-range parts ($\bm{F}_i^h$ and $\bm{T}_i^h$) are calculated using the direct forcing method~\cite{leonardi2015,pradipto2020b}, while the lubrication force $\bm{F}_i^{\rm lub}$ and torque $\bm{T}_i^{\rm lub}$ are expressed by pairwise interactions as $\bm{F}_{i}^{\rm lub} = \sum_{j \neq i} \bm{F}_{ij}^{\rm lub}$ and $\bm{T}_{i}^{\rm c} = \sum_{j \neq i} \bm{T}_{ij}^{\rm lub}$, respectively \cite{seto2013,mari2014,nguyen2002,pradipto2020}.
The explicit expressions of $\bm{F}_{ij}^{lub}$ and $\bm{T}_{ij}^{lub}$ can be found in Ref. \cite{pradipto2020}.

We adopt the linear spring-dashpot version of the DEM \cite{luding2008} for the contact interaction between particles, which involves both the normal and the tangential contact forces.
Note that we omit the dissipative part for the tangential contact force.
For the particle $i$, the contact force $\bm{F}^{c}_{i}$ and torque $\bm{T}^{c}_{i}$ are, respectively, written as $\bm{F}^{c}_{i} = \sum_{i \neq j} (\bm{F}^{\text{nor}}_{ij} + \bm{F}^{\text{tan}}_{ij} )$ and  $\bm{T}^{c}_{i} = \sum_{i \neq j} a_i \bm{n}_{ij} \times \bm{F}^{\text{tan}}_{ij}$, where $a_i$ is the radius of particle $i$.
The normal force is explicitly expressed as 
\begin{equation}
\bm{F}^{\text{nor}}_{ij} =( k_n \delta_{ij}^{n} -  \zeta^{(n)} u_{ij}^{(n)} ) \bm{n}_{ij}, 
\end{equation}
where $k_n$ is the spring constant, $\delta_{ij}^{n}$ is the normal overlap, $\bm{n}_{ij}$ is the normal unit vector between particles, $u_{ij}^{(n)}$ is the normal velocity difference of the contact point   $u_{ij}^{(n)} = u_i^{(n)}-u_j^{(n)}$, and $ \zeta^{(n)} = \sqrt{m_0 k_n}$ is the damping constant, where $m_0$ is the average mass of the suspended particles.
If the tangential contact force is smaller than a slip criterion, tangential contact force is represented as 
\begin{equation}
\tilde{\bm{F}}^{\text{tan}}_{ij} = k_t \delta_{ij}^{t} \bm{t}_{ij}, 
\end{equation}
where $k_t$, assumed to be $0.2k_n$, is the tangential spring constant, $\delta_{ij}^{t}$ is the tangential compression and $\bm{t}_{ij}$ is the tangential unit vector at the contact point between particles $i$ and $j$.
We adopt the Coulomb friction rules as
\begin{align}
|\bm{F}_{ij}^{\text{tan}}| &= \mu |\bm{F}_{ij}^{\text{nor}}| \quad \text{if } |\tilde{\bm{F}}^{\text{tan}}_{ij}| \geq \mu |\bm{F}^{\text{nor}}_{ij}|  \quad \text{(slip)}, \\
|\bm{F}_{ij}^{\text{tan}}| &= |\tilde{\bm{F}}^{\text{tan}}_{ij}| \quad  \text{if } |\tilde{\bm{F}}^{\text{tan}}_{ij}| \leq \mu |\bm{F}^{\text{nor}}_{ij}|  \quad \text{(stick)},
\end{align}
whereas $\delta_{ij}^{t}$ is updated each time with relative tangential velocity\cite{luding2008}.

Finally, $\bm{F}_i^r$ is the electrostatic repulsive force, also expressed by pairwise interactions as $\bm{F}_{i}^{\rm r} = \sum_{j \neq i} \bm{F}_{ij}^r$.
The explicit expression of $\bm{F}^{r}_{ij}$ is expressed by the Derjaguin-Landau-Verwey-Overbeek (DLVO) theory\cite{derjaguin1941,verwey1948,israelachvili2011} for the double layer electostatic force as
\begin{equation} 
\bm{F}^{r}_{ij} =  F_0\exp(-h/\lambda) \bm{n}_{ij}, 
\end{equation}
where $F_0=k_B T \lambda_B\hat{Z}^2(e^{a_{\text{min}}/\lambda}/(1+a_{\text{min}}/\lambda))^2/h^2$ with the charge number $\hat{Z}$, the Bjerrum length $\lambda_B$ and the Debye-H\"{u}ckel length $\lambda$. Note that $\lambda_B$ can be expressed as $\lambda_B=e^2/(4\pi \epsilon_0 \epsilon_r k_B T)$ where $e$, $\epsilon_0$, $\epsilon_r$, and $k_B$ are the elementary charge, the vacuum permittivity, the dielectric constant, and the Boltzmann constant, respectively\cite{israelachvili2011}.
Here, we adopt the Debye length $\lambda = 0.02a_{\text{min}}$.
Our simulation ignores the Brownian force.
Thus, the electrostatic repulsion force is important to prevent the suspended particles from clustering \cite{pradipto2020,mari2014}.

The impactor is a solid spherical object with the density $\rho_I = 4 \rho_f$.
The force and torque acting on the impactor are, respectively, given by 
\begin{equation}
	\bm{F}^{I} = \bm{F}^{I,h} + \bm{F}^{I,\rm lub} + \bm{F}^{I,c} + \bm{F}^{I,g},
\end{equation}
\begin{equation}	
	\bm{T}^{I} = \bm{T}^{I,h} +  \bm{T}^{I,c} +  \bm{T}^{I,\rm lub}.
\end{equation}
$\bm{F}^{I,g} = - m_I g \bm{\hat{z}}$ is the gravitational force acting on the impactor.
The contact force $\bm{F}^{I,c}$ and torque $\bm{T}^{I,c}$, which arise from the interactions with the suspended particles, are also calculated by the DEM.
The lubrication force $\bm{F}^{I,\rm lub}$ and torque $\bm{T}^{I,\rm lub}$ are also calculated in a similar manner as used in suspended particles.

The long-range hydrodynamic force $\bm{F}^{I,h}$ and torque $\bm{T}^{I,h}$ are calculated using the bounce-back rule which satisfies the no-slip boundary condition between the fluid and the surface of the impactor \cite{ladd1994a,ladd1994b}.
In the bounce-back rule the LBM discrete distribution function that streams from fluid nodes to the boundary nodes is reflected.
Then, the hydrodynamic force on each node is calculated from the momentum transferred in this reflection process.
In our implementation, the bounce-back rule is implemented by treating the surface of the impactor as boundary nodes.
An illustration of this bounce-back rule can be seen in Fig. \ref{fig:bb}.

\begin{figure}[htbp] 
	\includegraphics[width=0.7\linewidth]{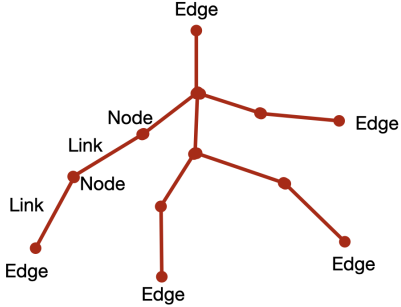}
	\caption{An illustration of the terminology in force chains. Lines represent the links, circles represent the nodes.} 
	\label{fig:edges}
\end{figure}

\section{\label{app:sol} Derivation and  analytical solution of the floating model}

\subsection{Derivation of $F_D^I$}

\begin{figure*}[htbp]	
	\subfloat[]{\label{fig:fc1}%
		\includegraphics[width=0.24\linewidth]{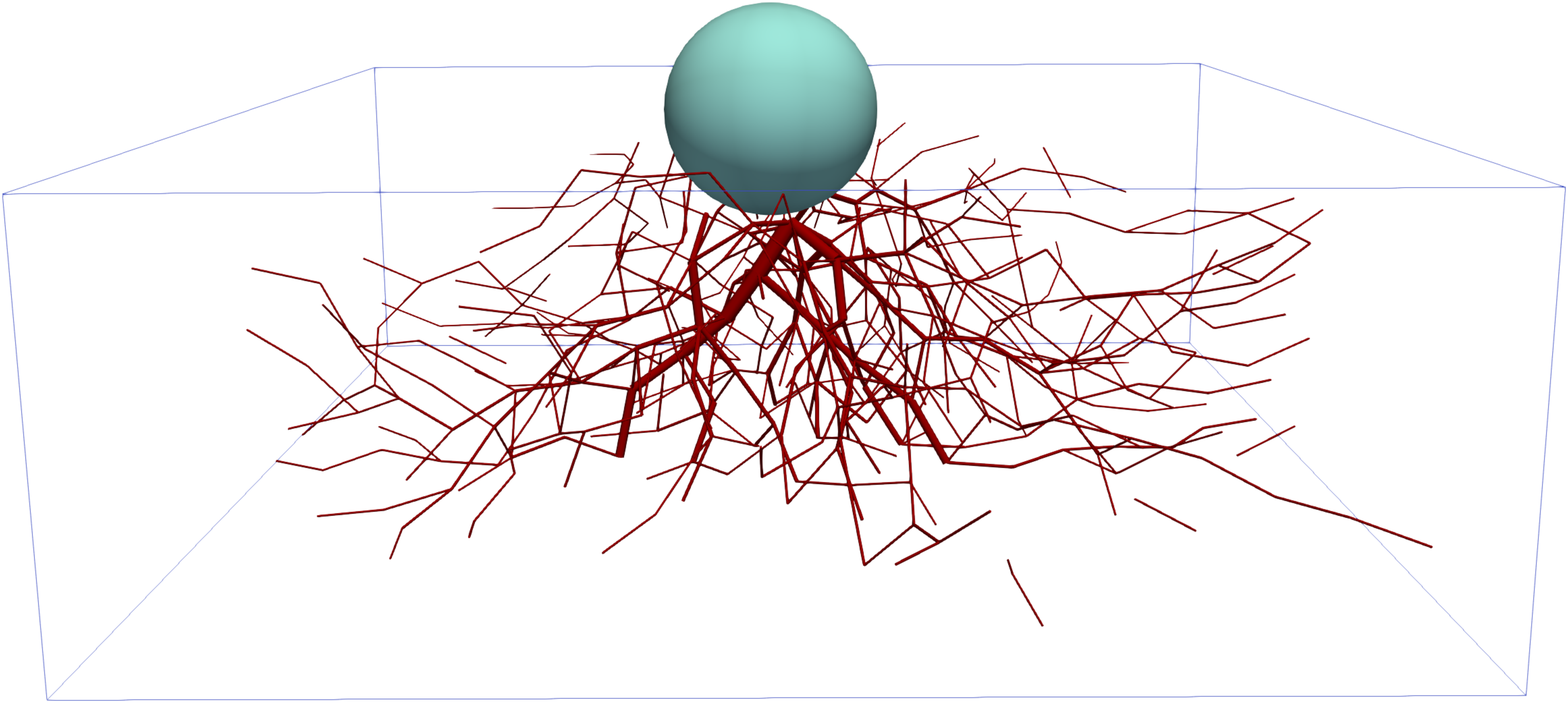}
	}
	\subfloat[]{\label{fig:fc2}%
		\includegraphics[width=0.24\linewidth]{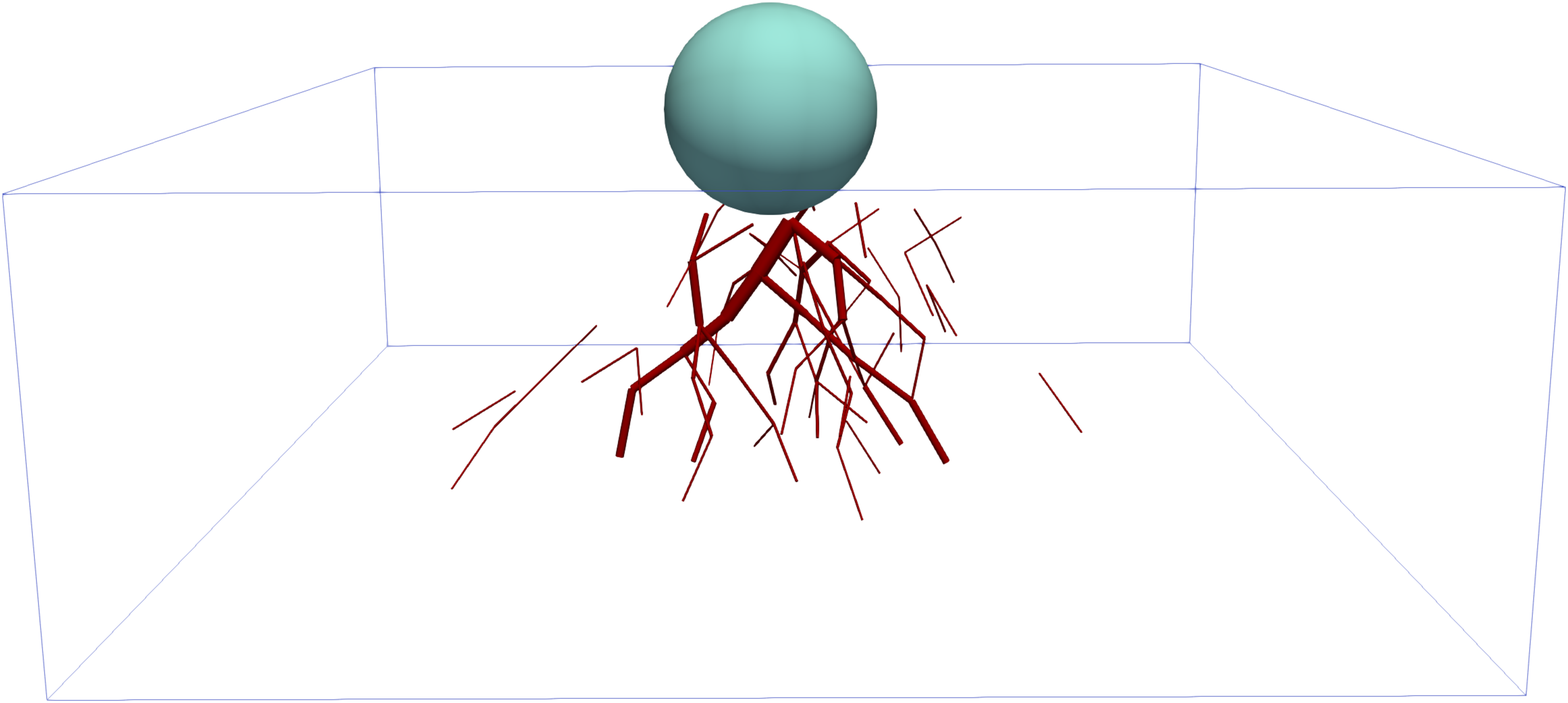}
	}
	\subfloat[]{\label{fig:fc3}%
		\includegraphics[width=0.24\linewidth]{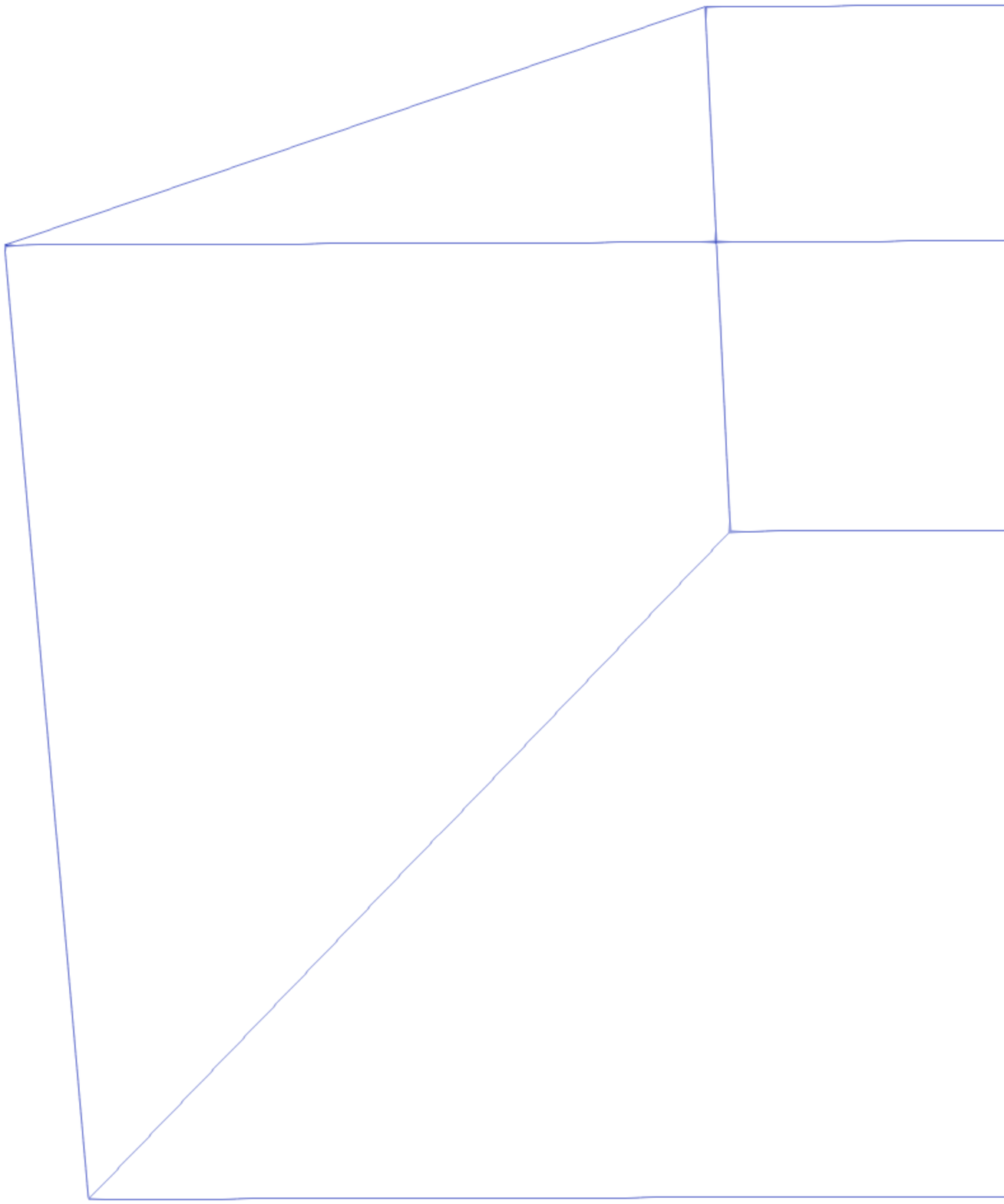}
	}
	\subfloat[]{\label{fig:fc4}%
		\includegraphics[width=0.24\linewidth]{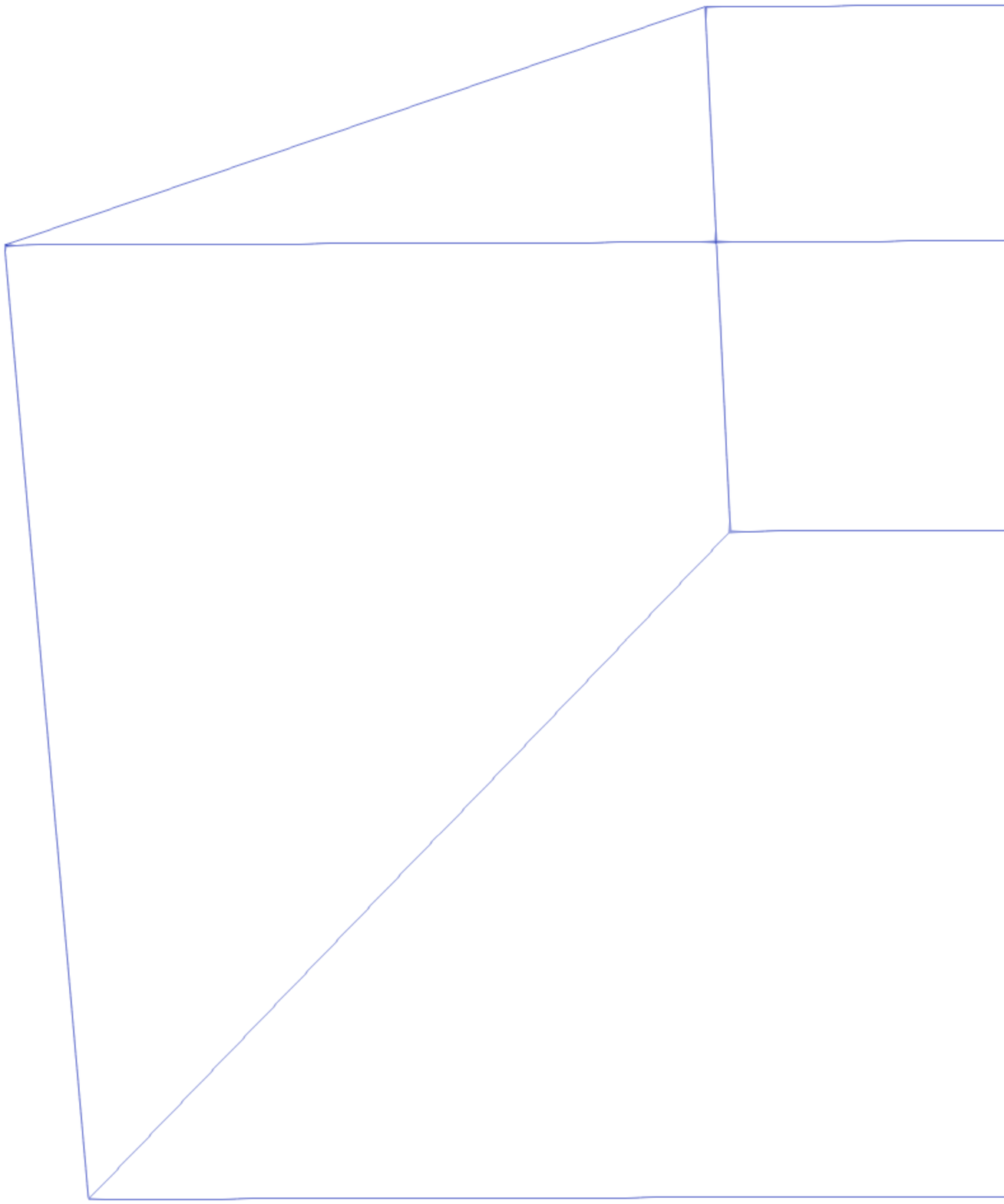}
	}
	\caption{Illustrations of the algorithm to determine $n(t)$ from force chains network (Multimedia view). (a)  An initial network pf force chains. (b) The remaining chains after the lateral chains are removed. (c) All connected components after removing all edges that do not touch the impactor or bottom plate, where blue and red lines represent the corresponding connected components that self-loops and chains between the impactor and bottom plate, respectively. (d) Percolated force chains from the impactor to the bottom plate.}
	\label{fig:fc}
\end{figure*}

The linear relationship between $|z|$ and the drag force in Fig. \ref{fig3} may be understood by the following simple model.
For $|z|<D_I$, the surface of the impactor is partially surrounded by the liquid and some parts of the surface are still in the air. 
Here, we assume that Stokes drag law can be used for the region surrounded by the liquid.
Thus, Stokes' drag force consists of two parts, the pressure drag $F_{D,p}^I$ and friction drag $F_{D,f}^I$ as $F^{I}_{D} = F^{I}_{D,p} + F^{I}_{D,f}$ \cite{batchelor2000}, 
\begin{align}
	F^{I}_{D,p} &=3\pi \eta_{\rm eff} a_I  \dot{z}_I  \int_0^{\theta_0}\cos^2\theta \sin\theta d\theta, \\
	F^{I}_{D,f}&=3\pi \eta_{\rm eff} a_I  \dot{z}_I \int_0^{\theta_0} \sin^3\theta d\theta \notag\\
	&=3\pi \eta_{\rm eff} a_I  \dot{z}_I(1-\cos \theta_0)- F^{I}_{D,p},
\end{align} 
where $a_I$ is the radius of the impactor satisfying $a_I=D_I/2$, $\eta_{\rm eff}$ is the  effective viscosity of the surrounding fluid, and $\theta_0$ is the separation angle between moving direction ($\theta=0$) and the line from the impactor center to the surface of the liquid (see Fig. \ref{fig:ilus2}). 
Note that the deepest position of the impactor satisfies the relation $|z|=a(1-\cos\theta_0)$.
Thus, one can reach Eq. \eqref{eq:drag}.

 \subsection{Analytical solution of Eq. \eqref{eq:eom}}
The dimensionless form of Eq. \eqref{eq:eom} with the aid of $a_I$ and $t_I=\sqrt{a_I/\tilde{g}}$ is written as
\begin{equation}
 \ddot{z}^{*}_I=  - 1 +\eta^{*} \dot{z}^{*}_I|z^{*}_I - 1|,
 \label{eq:dim}
\end{equation} 
where $z^{*}_I=z_I / a_I$, $t^{*} = t / t_I$, $\dot{z_I}^*=d z^{*}_I (t) /dt^{*}=-u_z^I/u^*$, $\ddot{z_I}^*=d^2 z^{*}_I (t) /dt^{*2}$, and $\eta^*=3\pi \eta_{\rm eff} a_I \sqrt{a_I/\tilde{g}}/m_I$.
Then, Eq. \eqref{eq:dim} can be solved exactly in terms of the Airy functions as
\begin{widetext}
	\begin{equation}
	z^{*}_I (t) =\frac{  \kappa[   -   \text{Ai}^{\prime}  (\Phi)\text{Bi}^{\prime} (\Theta)    + \text{Ai}^{\prime} (\Theta)  \text{Bi}^{\prime} (\Phi) ]  } { \gamma [  \text{Bi} (\Phi) \text{Ai}^{\prime} (\Theta)  -  \text{Ai} (\Phi)   \text{Bi}^{\prime} (\Theta)] },
	\label{eq:sol}
\end{equation}
\end{widetext}
where $\gamma =  (\eta^{*})^{2/3} $, $\kappa = 2^{2/3}$, $\Theta = u^{*}_0 \sqrt[3]{\eta^{*}/2} $, and $\Phi =(u_0^{*}+t^{*}) \sqrt[3]{\eta^{*}/2} $, where $u_0^*=u_0t_I/a_I$. 
Here, $\text{Ai} (x)$ is the Airy function of the first kind, which is defined as $\text{Ai} (x) =  \int_{0}^{\infty} \cos ( t^3/3 + xt )dt / \pi $, and $\text{Ai}^{\prime}(x)$ is its derivative. 
$\text{Bi}(x)$ is the Airy function of the second kind, which is defined as $\text{Bi} (x) =  \int_{0}^{\infty}[ \exp (- t^3/3 + xt ) +  \sin (-t^3/3 + xt) ]dt / \pi $, and $\text{Bi}^{\prime}(x)$ is its derivative. 
One can differentiate Eq. \eqref{eq:sol}  two times to get the expression for $\ddot{z}_I$ as
\begin{widetext}
	\begin{align}
	\ddot{z}^{*}_I = &\frac{\begin{matrix}   \bigg[  \kappa \gamma^{\frac{3}{2}}    \bigg(   u^{*}_0 -t^{*}\bigg) \bigg(\text{Ai}^{\prime} (\Theta)\text{Bi} (\Phi) - \text{Ai}(\Phi)\text{Bi}^{\prime} (\Theta) \bigg)^2    \bigg( \text{Ai}^{\prime} (\Theta) \text{Bi}^{\prime} (\Phi) - \text{Ai}^{\prime} (\Phi) \text{Bi}^{\prime} (\Theta) \bigg)  - \gamma \Lambda    \bigg]      \end {matrix}} { \bigg( \text{Ai}^{\prime} (\Theta) \text{Bi}(\Phi) -  \text{Ai} (\Phi) \text{Bi}^{\prime} (\Theta)\bigg)^{3}}, \notag\\
		\Lambda  = &\text{Ai}^{\prime} (\Theta)^{3} \text{Bi} (\Phi)^3 + 2 \text{Ai}^{\prime} (\Theta)^3\text{Bi}^{\prime} (\Phi)^3 - 3\text{Ai}(\Phi)\text{Ai}^{\prime} (\Theta)^2\text{Bi}^{\prime} (\Phi)^2\text{Bi}^{\prime} (\Theta) - 6 \text{Ai}^{\prime} (\Phi)\text{Ai}^{\prime} (\Theta)^2\text{Bi}^{\prime} (\Phi)^2\text{Bi}^{\prime} (\Theta) \notag\\
		& + 3 \text{Ai}(\Phi)^2\text{Ai}^{\prime} (\Theta)\text{Bi}(\Phi)\text{Bi}^{\prime} (\Theta)^2 + 6 \text{Ai}^{\prime} (\Phi)^2 \text{Ai}^{\prime} (\Theta)\text{Bi}^{\prime} (\Phi)\text{Bi}^{\prime} (\Theta)^2 + \text{Ai}(\Phi)^3\text{Bi}^{\prime} (\Theta)^3  - 2 \text{Ai}^{\prime} (\Phi)^3\text{Bi}^{\prime} (\Theta)^3. 			
		\label{eq:sol_f}
	\end{align}
\end{widetext}
To obtain the expression of $F_{\rm max}$ and $t_{\rm max}$, we adopt the short time expansion for Eq. \eqref{eq:sol_f}  since $F_{\rm max}$ appears in the region $t/t_g \ll 1$, 
Thus, up to third order, one can obtain
\begin{align}
	\ddot{z}^{*}_I = 1 &-  \eta^{*} u_0^{*2} t^{*} -  \frac{3 \eta^{*}  u^{*}_0 t^{*2}  }{2}  \notag\\
	&+ \bigg(  \frac{2 u^{*3}_0 (\eta^{*})^2 }{3}   -  \frac{ \eta^{*} }{2}   \bigg) t^{*3}   + \mathcal{O} \bigg( \left[ \frac{ t^{*}}{u^{*}_0}\right]^4\bigg)
	\label{eq:fser}
	\end{align}
Then, we differentiate Eq. \eqref{eq:fser} to obtain $	\dddot{z}^{*}_I$ as
\begin{align}
	\dddot{z}^{*}_I =  &-  \eta^{*} u_0^{*2}  -  3 \eta^{*}  u^{*}_0 t^{*}  \notag\\
&+ \bigg(  \frac{2 u^{*3}_0 (\eta^{*})^2 }{3}   -  \frac{ \eta^{*} }{2}   \bigg) t^{*2}   + \mathcal{O} \bigg( \left[ \frac{ t^{*}}{u^{*}_0}\right]^3\bigg)
		\label{eq:sol_dfdt}
	\end{align}
Then, for $	\dddot{z}^{*}_I  = 0$, one can solve the quadratic equation in Eq. \eqref{eq:sol_dfdt} for $t_{\rm max}$ as
\begin{equation}
		\frac{t_{\rm max}}{t_I} = \frac{ 3  u^{*}_0 + \sqrt{3   u_0^{*2} + 8   \eta^{*} u_0^{*5} } }{ 4  \eta^{*} u_0^{*3}  - 3  }.
	\label{eq:tm}
\end{equation}
For $ \eta_{\rm eff} u_0 \gg 1$, Eq. \eqref{eq:tm} reduces to
\begin{equation}
	\frac{t_{\rm max}}{t_I} = \frac{ u_0^{*-\frac{1}{2}}} {\sqrt{2}}.
			\label{eq:tm2}
\end{equation}
Thus, we confirm the exponent $\beta=-1/2$.
To obtain $F_{ \rm max}$, we plug Eq. \eqref{eq:tm2} into Eq. \eqref{eq:fser} and take the limit $ \eta_{\rm eff} u_0 \gg 1$.
Thus, we obtain
\begin{equation}
	\frac{F_{ \rm max}}{m_I \tilde{g}} = u_0^{*\frac{3}{2}} \sqrt{\frac{2 \eta^{*}}{9}}.
	\label{eq:fm}	
\end{equation}
Thus, we confirm the exponent $\alpha=3/2$ for large $u_0$.
From Eq. \eqref{eq:tm}, $t_{\rm max}$ diverges at $u_{0,c} = \sqrt[3]{3/ 4  \eta^{*} }$.
This result suggests the limitation of the short time approximation.

\section{\label{app:fc} Determination of $n(t)$ in the floating + force chain model}  

In this Appendix, we explain the algorithm to determine the connected force chains from the impactor to the bottom plate used in the floating + force chain model in Sec. \ref{chain}.
First of all, let us explain how we draw the force chains.
Note that force chains are defined as a collection of nodes and links representing the contacting suspended particles (see Fig. \ref{fig:edges}).
Thus, for each pair of contacting suspended particles, we draw a network in which a node represents the center of a contacting pair of particles and a link is a straight line connecting a pair of adjacent nodes.
The initial force chains can be seen in Fig. \ref{fig:fc1}.

The algorithm to determine $n(t)$  is as follows.
Since we are only interested in the force propagation in the vertical direction, we remove links that expand in the lateral directions (dangling chains).
For this purpose, we remove all links in which the height difference $|z_i - z_j|$ for a contacting pair of particles $i$ and $j$ is less than the smallest radius of the suspended particles $a_{\rm min}$.
The corresponding network after the removal of lateral chains can be seen in Fig. \ref{fig:fc2}.

Our goal is to determine connected networks from the impactor to the bottom plate.
Thus, we remove all links to reach the edges of the force chains which do not touch the bottom plate nor the impactor.
Once such links are removed, the leftover chains create new links at the edges of the force chain. 
We repeat these labeling and removal processes until there are no edges of dangling chains except for the edges which touch the bottom plate or the impactor.
Then, we label each connected component (the blue and red connected components in Fig. \ref{fig:fc3}).

\begin{figure}[htbp] 
	\includegraphics[width=0.8\linewidth]{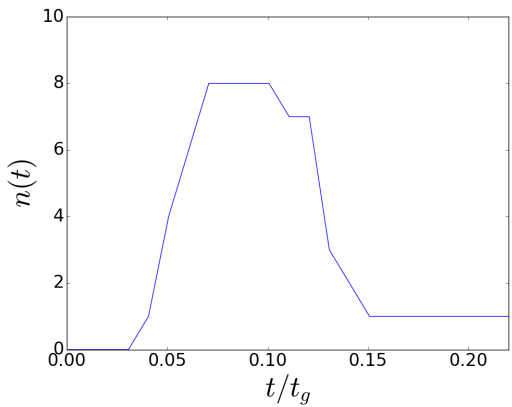}
	\caption{A plot of the number of connected force chains from the impactor to the bottom plate $n(t)$ against time for $\phi=0.53,$ $ W=D=6 D_I$, and $H=2D_I$ with $u_0 = 2.6 u^{*}$.}
	\label{fig6}
\end{figure}

Note that the connected components do not need to be percolated from the impactor to the bottom plate to survive in our algorithm at this stage due to the existence of connected edges which form a self-loop (blue connected components in Fig. \ref{fig:fc3}). 
Therefore, we need to examine whether each connected component is percolated or not.
Then, we remove non-percolated connected components (blue component) while keeping the percolated connected component (red component) as shown in Fig. \ref{fig:fc4}.
Finally, we evaluate $n(t)$ by the number of links that touch the bottom plate.
The above processes are illustrated in Fig. \ref{fig:fc} (Multimedia view).
The obtained $n(t)$ for $\phi=0.53,$ $ W=D=6 D_I$, and $H=2D_I$ with $u_0 = 2.6 u^{*}$ against time is plotted in Fig. \ref{fig6}.

\section{\label{app:dilute} Dependence on volume fraction of the suspensions.}

\begin{figure}[htbp] 
	\includegraphics[width=0.7\linewidth]{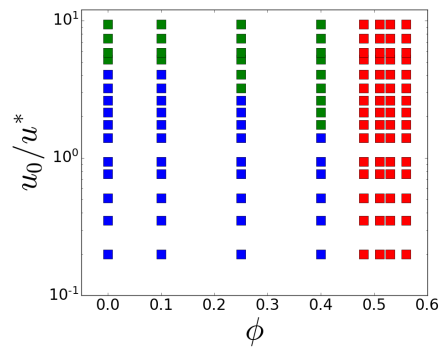}
	\caption{A phase diagram showing whether the impactor has $F_{\rm max}$ as a function of the volume fraction $\phi$ and the impact speed $u_0$ for $W=D=6D_I$ and $H=3D_I$. Red squares represent set of parameters where the relations among $u_0$, $F_{\rm max}$, and $t_{\rm max}$ can be explained by Eq. \eqref{eq:eom}. Green squares are points where $F_{\rm max}$ exist but Eq. \eqref{eq:eom} fails. Blue squares are where $F_{\rm max}$ does not even exist.} 
	\label{fig:phase}
\end{figure}

\begin{figure}[htbp]
	\subfloat[]{\label{fig:fdfast}%
		\includegraphics[width=0.7\linewidth]{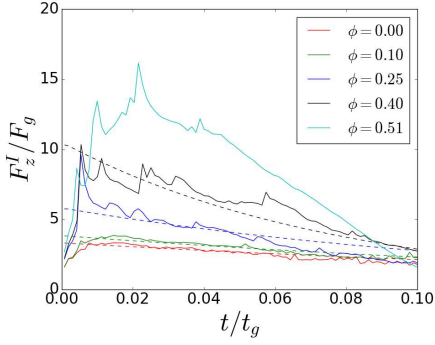}
	}
	
	\subfloat[]{\label{fig:fdslow}%
		\includegraphics[width=0.7\linewidth]{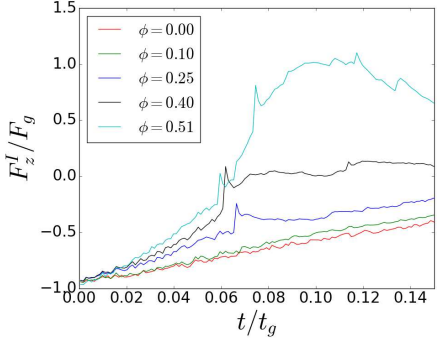}
	}
	\caption{Plots of forces exerted on the impactor against time for various volume fractions for $W=D=6D_I$ and $H=3D_I$ for (a) $u_0/u^{*} = 5.84$ (Dashed lines represent the solutions of Eq. \eqref{eq:stokes}) and (b) $u_0/u^{*} = 0.93$.  }
	\label{fig:fdilute}
\end{figure}

\begin{figure}[htbp] 
	\includegraphics[width=0.7\linewidth]{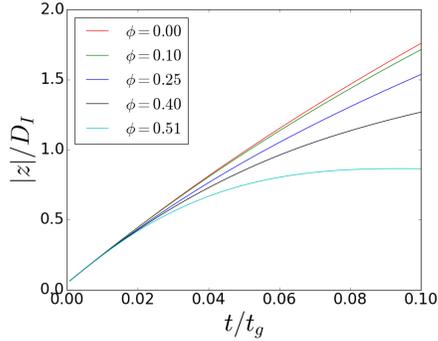}
	\caption{Plots of the deepest points of the impactor scaled by the diameter of the impactor for $u_0/u^{*} = 5.84$ for $W=D=6D_I$ and $H=3D_I$. }
	\label{fig:zdilute}
\end{figure}

\begin{figure}[htbp] 
	\includegraphics[width=0.7\linewidth]{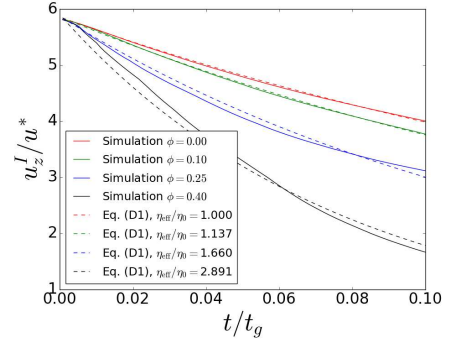}
	\caption{Plots of the velocities of the impactor for low volume fractions and for $W=D=6D_I$ and $H=3D_I$ alongside with the solutions of Eq. \eqref{eq:stokes}.}
	\label{fig:udilute}
\end{figure}

In this Appendix, we have examined whether the relations among $u_0$, $F_{\rm max}$, and $t_{\rm max}$ only exist in the impact process in dense suspensions, though the rebound only exists in dense suspensions.
We summarize the dependence on the volume fraction in the phase diagram in Fig. \ref{fig:phase}.
Our simulation indicates that $F_{\rm max}$ only exists in all range of $u_0$ when $\phi \geq 0.48$, while $F_{\rm max}$ does not exist for dilute suspensions except for very high $u_0$.
This result clarifies the role of suspensions in which our analysis in the main text is only valid for dense suspensions. 

In Fig. \ref{fig:fdilute}, we plot the force exerted on the impactor against time to clarify the difference between dense and dilute cases. 
For high $u_0$ (Fig. \ref{fig:fdfast}), $t_{\rm max}$ for the dilute case emerges earlier than that in the denser case.
In addition, dilute cases have smaller $F_{\rm max}$.
Such differences occur since the origin of $F_{\rm max}$ in a dilute case is different from that in the dense case.
In dense situations, the dominant contribution is from the contact force between the impactor and the suspended particles, while for the dilute situations, the dominant contribution is from the hydrodynamic force exerted on the impactor \cite{pradipto2020b}.
For low $u_0$ (Fig. \ref{fig:fdslow}), one can see that $F_{\rm max}$ only exists in dense situation.
Since the acceleration due to the gravity is dominant for low $u_0$, the sufficient drag resistance to compete with the gravity forces only exists for dense suspensions.

In Fig. \ref{fig:zdilute}, we plot the deepest point of the impactor $|z|$ (see Fig. \ref{fig:ilus2}) scaled by the impactor diameter $D_I$ against time.
Here, one can see that the impactor sinks right after the impact in dilute suspensions, while the impactor can keep its position near the surface for dense suspensions.
The behavior in which the impactor stays for a while near the surface of the suspension is a characteristic of dense suspensions under impact.
Thus, the floating model (Eq. \eqref{eq:eom}) cannot be used for dilute situations since the floating model assumes that the impactor is partially surrounded by fluid.
We also summarize the region where Eq. \eqref{eq:eom} is applicable in Fig. \ref{fig:phase}.
When the impactor is completely sink, the second term on the r.h.s. of Eq. \eqref{eq:eom} should be replaced by the Stokesian drag as
\begin{equation}
	m_I \frac{d^2 z_I}{dt^2}=- m_I \tilde{g} + 3\pi \eta_{\rm eff} a_I  \dot{z}_I.
	\label{eq:stokes}
\end{equation}
In Fig. \ref{fig:udilute}, we plot the impactor velocity against time alongside the solutions of Eq. \eqref{eq:stokes} for the dilute cases.
Note that Eq. \eqref{eq:eom} cannot describe even the behavior of the impactor velocity in dilute cases.

As expected, the apparent viscosity $\eta_{\rm eff}/\eta_0$, where $\eta_0$ is the solvent viscosity, becomes larger as the volume fraction increases.
Although the agreement between the solution of Eq. \eqref{eq:stokes} and our simulation is remarkable for $u_z^I$ (see Fig. \ref{fig:udilute}),  Eq. \eqref{eq:stokes} cannot capture $F_{\rm max}$ in dilute suspensions (see the dashed lines in Fig. \ref{fig:fdfast}).
This is because there is no competition between time increasing and time decreasing contributions in Eq. \eqref{eq:stokes}.
Although one may extend the studies on the impact process on water (without suspended particles) alone to dilute suspensions \cite{moghisi1981}, such a problem is beyond the scope of this paper.

\bibliography{references}

\end{document}